\documentclass[lettersize,journal]{IEEEtran}
\usepackage{amsmath,amsfonts}
\usepackage{algorithmic}
\usepackage{algorithm}
\usepackage{array}
\usepackage[caption=false,font=scriptsize,labelfont=rm,textfont=rm]{subfig}
\usepackage{textcomp}
\usepackage{stfloats}
\usepackage{url}
\usepackage{verbatim}
\usepackage{graphicx}
\usepackage{cite}
\usepackage{bm}			
\usepackage{amssymb}		
\usepackage{amsthm}
\newtheorem{theorem}{Theorem}

\newtheorem{corollary}{Corollary}

\allowdisplaybreaks

\usepackage[colorlinks=true,citecolor=blue,linkcolor=blue,urlcolor=blue]{hyperref}

\begin{document}

\title{ On the Secrecy Performance of Pinching-Antenna Systems}

\author{Nianzu Li,~Weidong Mei,~\IEEEmembership{Member,~IEEE},~Lipeng Zhu,~\IEEEmembership{Member,~IEEE},~Peiran Wu,~\IEEEmembership{Member,~IEEE},\\~and~Boyu Ning,~\IEEEmembership{Member,~IEEE} \vspace{-2em}
\thanks{Nianzu Li and Peiran Wu are with the School of Electronics and Information Technology, Sun Yat-sen University, Guangzhou 510006, China (e-mail: linz5@mail2.sysu.edu.cn; wupr3@mail.sysu.edu.cn).}

\thanks{Weidong Mei and Boyu Ning are with the National Key Laboratory of Wireless Communications, University of Electronic Science and Technology of China, Chengdu 611731, China (e-mail: wmei@uestc.edu.cn; boydning@outlook.com).}

\thanks{Lipeng Zhu is with the Department of Electrical and Computer Engineering, National University of Singapore, Singapore 117583 (e-mail: zhulp@nus.edu.sg).}
	
}

\markboth{Journal of \LaTeX\ Class Files,~Vol.~18, No.~9, September~2020}%
{Shell \MakeLowercase{\textit{et al.}}: A Sample Article Using IEEEtran.cls for IEEE Journals}


\maketitle

\begin{abstract}
	Pinching-antenna systems have recently gained significant attention as a novel reconfigurable-antenna technology due to its exceptional capability of mitigating signal-propagation path loss. In this letter, we investigate the secrecy performance of a pinching-antenna system in the presence of an eavesdropper. In particular, we derive an approximate expression of the system's secrecy outage probability (SOP) with respect to the random locations of the legitimate user and eavesdropper and analyze its asymptotic behavior. Moreover, we derive a constant performance lower bound on the SOP of the considered system, i.e., $\frac{2\pi-1}{24}$, which is significantly lower than that of conventional fixed-position antenna systems, i.e., $0.5$. Finally, simulation results are provided to validate the correctness of our analytical results.

\end{abstract}


\section{Introduction}
\IEEEPARstart{W}{ith} the rapid proliferation of various wireless applications, such as high-quality video streaming, telemedicine, Internet of Things, the upcoming sixth generation (6G) wireless communication networks are poised to meet the versatile demand of higher capacity, lower latency, and enhanced reliability. However, signal-propagation path loss, a long-standing challenge in wireless communications, poses a critical hurdle to realizing the above vision. In this context, various reconfigurable antenna systems have been developed in recent years as a viable technology to break through this obstacle \cite{ref47}, \cite{ref62},\cite{ref63},\cite{ref64}. In particular, pinching antennas utilize a long dielectric waveguide as signal transmission mediums and enable flexible antenna position activation at any location along the waveguide by deploying small dielectric particles on it. This innovative antenna architecture can reduce the signal-propagation distance between the antenna's radiating points and users, thereby offering great potential to mitigate the large-scale path loss\cite{ref49},\cite{ref60}.

Motivated by such a promising benefit, many recent works have optimized the performance of pinching-antenna systems in various wireless communication scenarios. In \cite{ref50}, the authors investigated a rate maximization problem in a downlink pinching-antenna system, aiming to maximize a single user's achievable rate by optimizing the antenna location activation. In \cite{ref61}, the authors studied the performance optimization of a downlink multi-user pinching-antenna system, with a goal to maximize the system's energy efficiency via joint antenna positioning and resource allocation. In \cite{ref52}, the authors proposed to apply pinching antennas in integrated sensing and communications (ISAC) systems. Their results show that the ISAC performance can be remarkably improved by leveraging the capability of pinching antennas to establish strong line-of-sight transmission. Besides, exploring pinching antennas for enhancing physical layer security was considered in \cite{ref57}, \cite{ref58}. However, most of the existing works focus their attention on developing antenna-positioning optimization algorithms, while an in-depth theoretical analysis of pinching-antenna systems still remains largely unexplored.


\begin{figure}[t]
	\centering
	\includegraphics[width=0.45\textwidth]{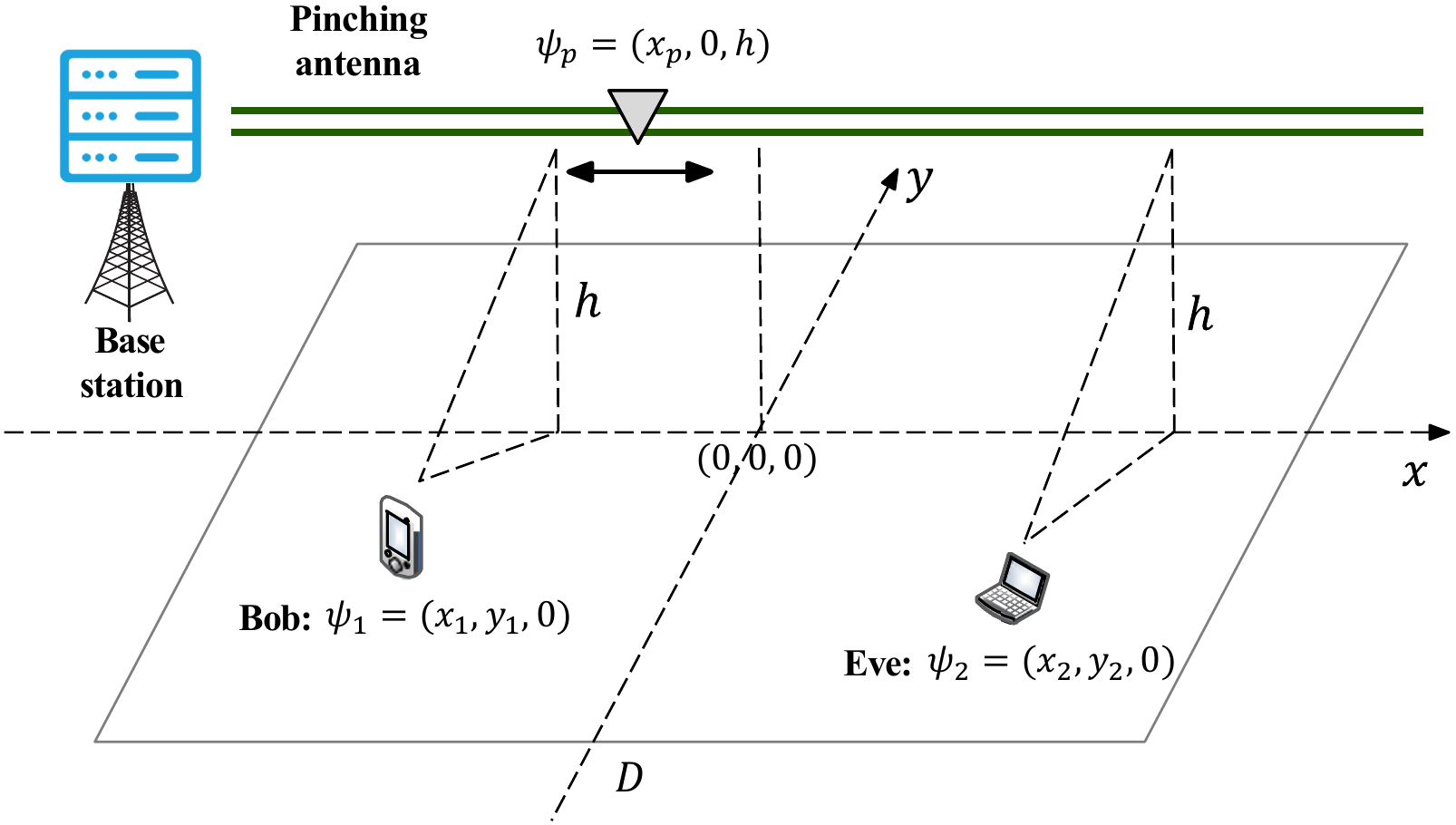}
	\caption{Secure pinching-antenna system.}
	\label{system_model}
	\vspace{-0.5cm}
\end{figure}
To fill in this gap, this letter conducts performance analysis of a pinching antenna-enhanced secure communication system. As shown in Fig. \ref{system_model}, we consider the downlink transmission from a pinching-antenna base station (BS) or access point (AP) to a single-antenna legitimate user (Bob) in the presence of a single-antenna eavesdropper (Eve). Our goal is to analyze the system's secrecy outage probability (SOP) with respect to the random positions of the legitimate user and eavesdropper. First, by leveraging Gaussian-Chebyshev quadrature, we obtain an analytical approximate expression of the SOP, followed by which its asymptotic behavior is given. Then, to gain more insights, we derive constant lower bounds on the SOP of our considered pinching-antenna system as well as conventional fixed-position antenna (FPA) systems, showing that the former is significantly lower than the latter ($\frac{2\pi-1}{24}\approx0.2201$ versus $0.5$). Numerical results are finally provided to validate the correctness of our analytical results.

\textit{Notations}: Boldface lowercase and uppercase letters denote vectors and matrices, respectively. $\|\mathbf{x}\|$ denotes the Euclidean norm of vector $\mathbf{x}$. $x\sim\mathcal{U}[a,b]$ represents that $x$ follows a uniform distribution over $[a,b]$.  $\mathrm{Pr}\left(\cdot\right)$ denotes the probability of an event. $\int_{a}^{b}f(x)dx$ denotes the definite integral of $f(x)$ from $a$ to $b$.

\section{System model}
As illustrated in Fig. \ref{system_model}, we consider the downlink of a secure communication system, which consists of a BS, a legitimate user (Bob), and an eavesdropper (Eve) aiming to intercept the information intended for Bob. Both Bob and Eve are equipped with a single FPA while the BS employs a pinching antenna on a dielectric waveguide for enhancing the system's secrecy performance. The waveguide is assumed to be installed parallel to the $x$-axis at a height $h$. Hence, the position of the pinching antenna can be denoted by $\bm{\psi}_{p}=(x_p,0,h)$. Besides, the positions of Bob and Eve are assumed to be uniformly distributed within the $x$-$y$ plane with a size of $D\times D$ and are denoted by $\bm{\psi}_{1}=(x_1,y_1,0)$ and $\bm{\psi}_{2}=(x_2,y_2,0)$, respectively. In this letter, we assume Eve is passive, such that the BS only has Bob's channel state information (CSI). Thus, the pinching antenna is activated at the position that is closest to Bob, i.e., $\tilde{\bm{\psi}}_{p}=(x_1,0,h)$.
\begin{figure*}[t!]
	\setcounter{equation}{6}
	\begin{align}
		\label{eq7}
		f_{\gamma_E}(z)=\begin{cases}
			\frac{\bar{\gamma}}{z^2}\left[\frac{\pi}{D^2}-\frac{2}{D^3}\sqrt{\frac{\bar{\gamma}}{z}-h^2}\right], &\frac{\bar{\gamma}}{h^2+\frac{D^2}{4}}\leq z \leq \frac{\bar{\gamma}}{h^2},\\
			\frac{2\bar{\gamma}}{z^2D^2}\left[\arcsin\big(\frac{D}{2\sqrt{\frac{\bar{\gamma}}{z}-h^2}}\big)-\frac{1}{2}\right], &\frac{\bar{\gamma}}{h^2+D^2}\leq z < \frac{\bar{\gamma}}{h^2+\frac{D^2}{4}},\\
			\frac{2\bar{\gamma}}{z^2D^3}\left[D\left(\arcsin\big(\frac{D}{2\sqrt{\frac{\bar{\gamma}}{z}-h^2}}\big)-\arccos\big(\frac{D}{\sqrt{\frac{\bar{\gamma}}{z}-h^2}}\big)\right)-\left(\frac{D}{2}-\sqrt{\frac{\bar{\gamma}}{z}-h^2-D^2}\right)\right], &\frac{\bar{\gamma}}{h^2+\frac{5D^2}{4}}\leq z < \frac{\bar{\gamma}}{h^2+D^2},\\
			0, &\mathrm{otherwise}.
		\end{cases}
	\end{align}
	\hrulefill
\end{figure*}

\subsection{Signal-to-Noise Ratio (SNR) at Bob}
Similarly as in \cite{ref47}, we adopt the spherical wave model to characterize the channel between the antenna activation point and Bob, which is given by
\begin{align}
	\setcounter{equation}{0}
	\label{eq1}
	h_B=\frac{\eta^{\frac{1}{2}}e^{-j\frac{2\pi}{\lambda}\|\tilde{\bm{\psi}}_{p}-\bm{\psi}_1\|}}{\|\tilde{\bm{\psi}}_{p}-\bm{\psi}_1\|},
\end{align}
where $\eta=\frac{\lambda^2}{16\pi^2}$, with $\lambda$ denoting the signal wavelength. Then, the received signal at Bob can be expressed as
\begin{align}
	\label{eq2}
	r_B=\sqrt{P_s}\frac{\eta^{\frac{1}{2}}e^{-j\frac{2\pi}{\lambda}\|\tilde{\bm{\psi}}_{p}-\bm{\psi}_1\|}}{\|\tilde{\bm{\psi}}_{p}-\bm{\psi}_1\|}e^{-j\phi}s+n_B,
\end{align}
where $P_s$ denotes the transmit power of the BS, $s$ denotes the transmitted signal from the BS with a unit power, and $n_B$ is the additive white Gaussian noise with zero mean and variance $\sigma^2$. Besides, $\phi=\frac{2\pi}{\lambda_g}\|\tilde{\bm{\psi}}_{p}-\bm{\psi}_0\|$ denotes the phase shift incurred by the signal propagation inside the waveguide, where $\bm{\psi}_0=(-\frac{D}{2},0,h)$ is the feed-point position of the transmitted signal and $\lambda_g=\frac{\lambda}{n_{\mathrm{eff}}}$ is the guided wavelength in the waveguide. 

Based on \eqref{eq2}, the SNR at Bob is expressed as
\begin{align}
	\label{eq3}
	\gamma_B=\frac{\eta P_s}{\|\tilde{\bm{\psi}}_{p}-\bm{\psi}_1\|^2\sigma^2}=\frac{\bar{\gamma}}{y_1^2+h^2},
\end{align}
where $\bar{\gamma}=\frac{\eta P_s}{\sigma^2}$ denotes the effective transmit SNR. Based on \cite[Proposition 2]{ref48}, the cumulative distribution function (CDF) of $\gamma_B$ with respect to the random position of Bob is given by
\begin{align}
	\label{eq4}
	F_{\gamma_B}(z)=\begin{cases}
		1, &z\geq\frac{\bar{\gamma}}{h^2},\\
		1-\frac{2}{D}\sqrt{\frac{\bar{\gamma}}{z}-h^2}, &\frac{\bar{\gamma}}{h^2+\frac{D^2}{4}}\leq z\leq \frac{\bar{\gamma}}{h^2},\\
		0, &z \leq \frac{\bar{\gamma}}{h^2+\frac{D^2}{4}}.
	\end{cases}
\end{align}

\subsection{SNR at Eve}
Similarly, the received signal at Eve can be expressed as
\begin{align}
	\label{eq5}
	r_E=\sqrt{P_s}\frac{\eta^{\frac{1}{2}}e^{-j\frac{2\pi}{\lambda}\|\tilde{\bm{\psi}}_{p}-\bm{\psi}_2\|}}{\|\tilde{\bm{\psi}}_{p}-\bm{\psi}_2\|}e^{-j\phi}s+n_E,
\end{align}
where $n_E$ is the additive white Gaussian noise with zero mean and variance $\sigma^2$. Then, the SNR at Eve can be obtained as
\begin{align}
	\label{eq6}
	\gamma_E=\frac{\eta P_s}{\|\tilde{\bm{\psi}}_{p}-\bm{\psi}_2\|^2\sigma^2}=\frac{\bar{\gamma}}{(x_1-x_2)^2+y_2^2+h^2}.
\end{align}
Since $x_1$, $x_2$, and $y_2$ are independent random variables with uniform distributions over $[-\frac{D}{2},\frac{D}{2}]$, the probability density function (PDF) of $\gamma_E$ can be obtained as \eqref{eq7}, shown at the top of this page. The details are provided in Appendix A. 

\subsection{Performance Metric}
In this letter, we adopt the SOP as the performance metric of the considered secure pinching-antenna system. Specifically, we assume that the BS transmits signals to Bob at a constant target secrecy rate, $R_{th},$ and a secure transmission can only be guaranteed if $R_{th}$ is lower than the system's instantaneous achievable secrecy rate. The SOP is, therefore, defined as
\begin{align}
	\setcounter{equation}{7}
	\label{eq8}
	\mathcal{P}_{so}=\mathrm{Pr}\left(\log_2\left(1+\gamma_B\right)-\log_2\left(1+\gamma_E\right)\leq R_{th}\right),
\end{align}
which characterizes the probability that a secure transmission cannot be realized.

\textit{Remark 1:} Note that since we consider a passive eavesdropper scenario in which the BS does not have the instantaneous CSI knowledge of Eve, the BS only transmits information at a fixed secrecy rate. In this case, the SOP is adopted as a typical metric to evaluate the system's secrecy performance \cite{ref59}. The active eavesdropping case in which the CSI of both Bob and Eve is available at the BS can be pursued in our future work.


\section{Secrecy Outage Probability Analysis}
In this section, we derive the analytical expression of the system's SOP as well as its asymptotic  behavior and drive insights by comparing with conventional FPA systems.

\subsection{SOP Analysis}
According to \eqref{eq8}, the SOP can be equivalently expressed as
\begin{align}
	\label{eq9}
	\mathcal{P}_{so}&=\mathrm{Pr}\left(\frac{1+\gamma_B}{1+\gamma_E}\leq C_{th}\right)\notag\\
	&=\mathrm{Pr}\left(1+\gamma_B-C_{th}\leq C_{th}\gamma_E\right),
\end{align}
where $C_{th}=2^{R_{th}}$. By utilizing the CDF of $\gamma_B$ and the PDF of $\gamma_E$ in \eqref{eq4} and \eqref{eq7}, the SOP can be further expanded as
\begin{align}
	\label{eq10}
	\mathcal{P}_{so}&=\int_{0}^{+\infty}f_{\gamma_E}(t)\mathrm{Pr}\left(\gamma_B\leq C_{th}t+C_{th}-1\right) dt \notag\\
	 &=\int_{\frac{\bar{\gamma}}{h^2+\frac{5D^2}{4}}}^{\frac{\bar{\gamma}}{h^2}}f_{\gamma_E}(t)F_{\gamma_B}\left(C_{th}t+C_{th}-1\right) dt,
\end{align}
where the integral range is due to the fact that $\gamma_E$ in \eqref{eq6} satisfies  $\frac{\bar{\gamma}}{h^2+\frac{5D^2}{4}}\leq \gamma_E\leq\frac{\bar{\gamma}}{h^2}$. The upper bound is achieved when $x_1-x_2=y_2=0$, while the lower bound is achieved when $x_1-x_2=\pm D$ and $y_2=\pm \frac{D}{2}$.
However, due to the cumbersome integration, it is still challenging to analyze the SOP based on \eqref{eq10}. Therefore, we present the following theorem, which provides an approximate closed-form expression of the system's SOP.
\begin{theorem}
	For the considered secure pinching-antenna system, the SOP can be approximated by
	\begin{align}
		\label{eq11}
		\mathcal{P}_{so}\approx\frac{\pi}{N}\sum_{n=1}^{N}& \sqrt{1-\cos^2\left(\frac{2n-1}{N}\pi\right)}\Gamma\left(\cos\left(\frac{2n-1}{N}\pi\right)\right)\notag\\
		&\times \left( \frac{\bar{\gamma}}{2h^2}-\frac{\bar{\gamma}}{2h^2+\frac{5D^2}{2}} \right),
	\end{align}
	where $\Gamma(x)=f_{\gamma_E}\Big(\Big( \frac{\bar{\gamma}}{2h^2}-\frac{\bar{\gamma}}{2h^2+\frac{5D^2}{2}} \Big)x+\frac{\bar{\gamma}}{2h^2}+\frac{\bar{\gamma}}{2h^2+\frac{5D^2}{2}}\Big)\cdot F_{\gamma_B}\Big(C_{th}\Big(\Big( \frac{\bar{\gamma}}{2h^2}-\frac{\bar{\gamma}}{2h^2+\frac{5D^2}{2}} \Big)x+\frac{\bar{\gamma}}{2h^2}+\frac{\bar{\gamma}}{2h^2+\frac{5D^2}{2}}\Big)+C_{th}-1\Big)$ and $N$ is a tunable parameter to balance the complexity and accuracy.
\end{theorem}

\textit{Proof:} First, let us define $u=\frac{2t-\left(\frac{\bar{\gamma}}{h^2}+\frac{\bar{\gamma}}{h^2+\frac{5D^2}{4}}\right)}{\frac{\bar{\gamma}}{h^2}-\frac{\bar{\gamma}}{h^2+\frac{5D^2}{4}}}$. Then, the integral interval in \eqref{eq10} can be transformed into $[-1,1]$, i.e.,
\begin{align}
	\label{eq12}
	\mathcal{P}_{so}
	=&\int_{-1}^{1}f_{\gamma_E}\left(g(u)\right)F_{\gamma_B}\left(C_{th}g(u)+C_{th}-1\right) \notag\\
	&\times \left( \frac{\bar{\gamma}}{2h^2}-\frac{\bar{\gamma}}{2h^2+\frac{5D^2}{2}} \right) du,
\end{align} 
where $g(u)=\left( \frac{\bar{\gamma}}{2h^2}-\frac{\bar{\gamma}}{2h^2+\frac{5D^2}{2}} \right)u+\frac{\bar{\gamma}}{2h^2}+\frac{\bar{\gamma}}{2h^2+\frac{5D^2}{2}}$.
Next, by applying $N$-point Gaussian-Chebyshev quadrature \cite{ref55}, we can derive the approximated SOP in \eqref{eq11}. $\hfill\square$


\subsection{Asymptotic Behavior of SOP} 
To gain further insights, we analyze the asymptotic behavior of the system's SOP with respect to the transmit power of the BS. First, for a sufficiently large $P_s$, i.e., $P_s\rightarrow\infty$, we have $\bar{\gamma}=\frac{\eta P_s}{\sigma^2}\rightarrow\infty$. Therefore, the terms $\log_2(1+\gamma_B)$ and $\log_2(1+\gamma_B)$ in \eqref{eq8} can be approximated by
\begin{align}
	\label{eq13}
	\log_2(1+\gamma_B)\rightarrow\log_2\left(\frac{\bar{\gamma}}{y_1^2+h^2}\right),
\end{align}
and
\begin{align}
	\label{eq14}
	\log_2(1+\gamma_E)\rightarrow\log_2\left(\frac{\bar{\gamma}}{(x_1-x_2)^2+y_2^2+h^2}\right),
\end{align}
respectively. Then, the asymptotic SOP can be obtained as
\begin{align}
	\label{eq15}
	\mathcal{P}_{so}&\approx\mathrm{Pr}\left(\log_2\left(\frac{(x_1-x_2)^2+y_2^2+h^2}{y_1^2+h^2}\right)\leq R_{th}\right)\notag\\
	&\approx\mathrm{Pr}\left(\frac{(x_1-x_2)^2+y_2^2+h^2}{y_1^2+h^2}\leq C_{th}\right).
\end{align}

Note that \eqref{eq15} unveils that with a given target secrecy rate $R_{th}$, the system's SOP will converge to a constant value as the BS's transmit power grows, which yields the following corollary.

\begin{corollary}
	At a high transmit SNR regime, the asymptotic SOP of the considered secure pinching-antenna system can be obtained as a constant value, i.e.,
	\begin{align}
		\label{eq16}
		\mathcal{P}_{so}\approx\frac{2}{D}\int_{0}^{D/2} F_{\chi}\left(C_{th}(t^2+h^2)-h^2\right) dt,
	\end{align}
	where $F_{\chi}(\cdot)$ denotes the CDF of $\chi=(x_1-x_2)^2+y_2^2$, which can be obtained similarly as the derivation of \eqref{eq7} presented in Appendix A and thus omitted here for brevity.
\end{corollary}

\subsection{Lower Bound on SOP}
Next, to demonstrate the advantage of the pinching antenna in enhancing system's security, we derive a theoretical lower bound on the SOP for the considered pinching-antenna system and compare it with that for traditional FPA systems. 
Specifically, for FPA systems, we assume that the BS's antenna is deployed at the center of the deployment region size with a height $h$, i.e., $\hat{\bm{\psi}}_0=(0,0,h)$. As a result, the corresponding SNR at Bob and Eve is given by
\begin{align}
	\label{eq17}
	\hat{\gamma}_B=\frac{\eta P_s}{\|\hat{\bm{\psi}}_0-\bm{\psi}_1\|^2\sigma^2}=\frac{\bar{\gamma}}{x_1^2+y_1^2+h^2},
\end{align}
and
\begin{align}
	\label{eq18}
	\hat{\gamma}_E=\frac{\eta P_s}{\|\hat{\bm{\psi}}_0-\bm{\psi}_2\|^2\sigma^2}=\frac{\bar{\gamma}}{x_2^2+y_2^2+h^2},
\end{align}
respectively. Hence, the SOP of the conventional FPA system is expressed as
\begin{align}
	\label{eq19}
	\hat{\mathcal{P}}_{so}=\mathrm{Pr}\left(\log_2\left(1+\hat{\gamma}_B\right)-\log_2\left(1+\hat{\gamma}_E\right)\leq R_{th}\right).
\end{align}

\begin{theorem}
	For the considered secure pinching-antenna system, the SOP is lower bounded by
	$\mathcal{P}_{so,low}=\frac{2\pi-1}{24}\approx0.2201$; while for conventional FPA systems, the SOP is lower bounded by $\hat{\mathcal{P}}_{so,low}=0.5$.
\end{theorem}

\textit{Proof:} Please refer to Appendix B. $\hfill\square$

Theorem 2 demonstrates that regardless of Bob and Eve's distribution region size and BS transmit power, the considered pinching-antenna system can achieve a much lower bound on SOP than its FPA counterpart. This is because for FPA systems, the instantaneous achievable secrecy rate is dominated by the distances between the BS and Bob/Eve due to the fixed antenna layout. Thus, the inherent random distributions of Bob and Eve will result in a high SOP. In contrast, for pinching-antenna systems, the position of the antenna can be flexibly adjusted to reduce the propagation distance between the BS and Bob, thereby enhancing the SOP performance.


\section{Numerical results}
In this section, simulation results are provided to examine the accuracy and validity of our derived analytical results. The simulation parameters are set as follows unless otherwise stated. The waveguide is deployed at a height of $h=3$ m. The carrier frequency is set to $\lambda=28$ GHz and the effective refractive index is $n_{\mathrm{eff}}=1.4$. Besides, the received noise power is set to $\sigma^2=-80$ dBm. The number of sampling points for Gaussian-Chebyshev quadrature is set to $N=100$. In all figures, we present the SOP derived from Monte Carlo (MC) simulations with $10^{5}$ channel trials and compare it with our analytical results.

\begin{figure}[!t]
	\centering
	\subfloat[]{
		\includegraphics[width=0.45\textwidth]{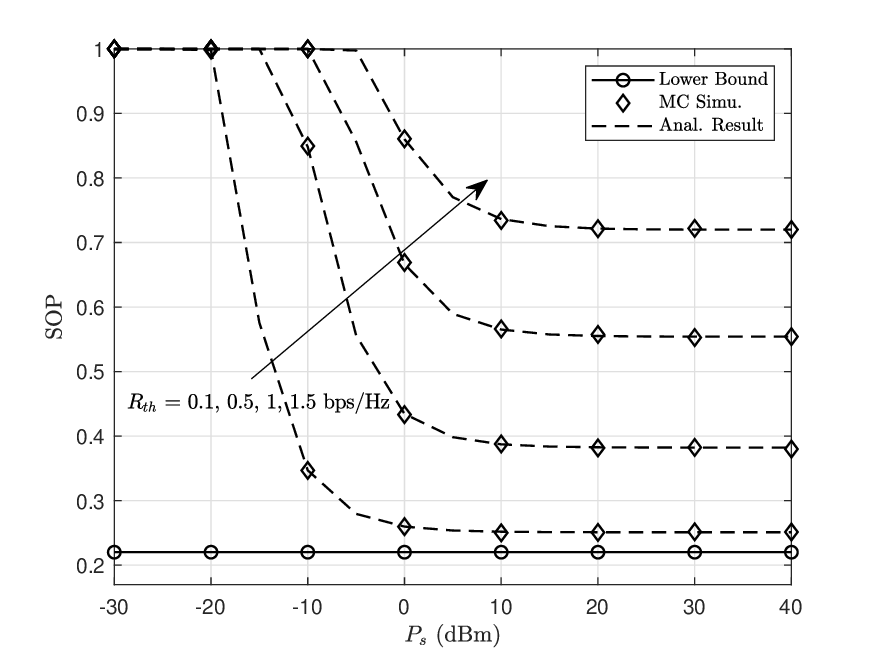} 
	}
	\vspace{-0.1cm}
	\subfloat[]{
		\includegraphics[width=0.45\textwidth]{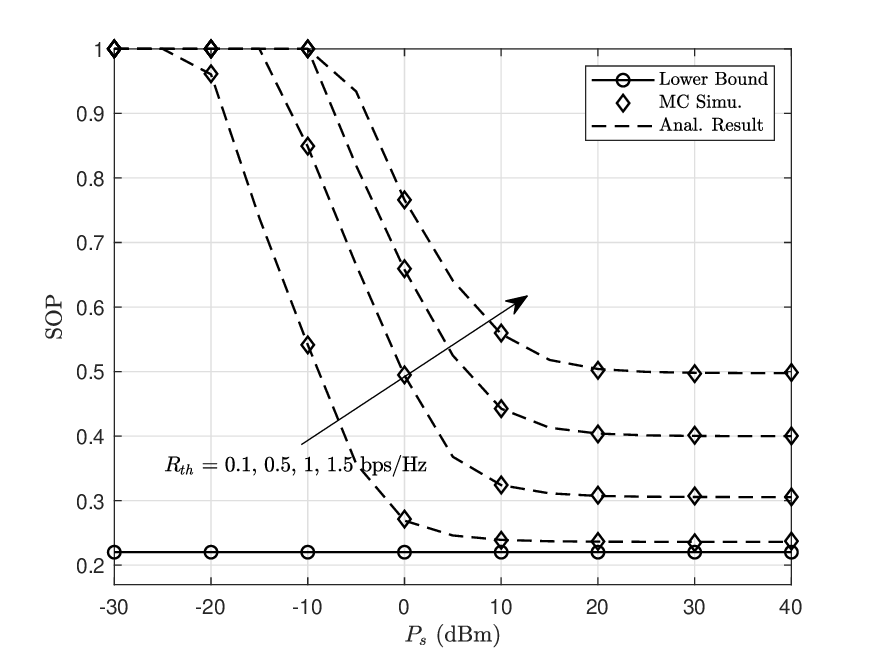} 
	}
	\caption{Secrecy outage probability of the considered secure pinching-antenna system versus the transmit power (in dBm) for different values of $R_{th}$ with: (a) $D=10$ m, (b) $D=30$ m.}
	\label{simluations}
	\vspace{-0.3cm}
\end{figure}
In Fig. \ref{simluations}, we plot the SOP versus the transmit power of the BS based on MC simulations and our analytical expressions in Theorems 1 and 2, with different values of the deployment region size, i.e., $D=10$ and $30$ m. As can be observed, the SOP derived from MC simulations matches well with \eqref{eq11} under any given $P_s$, $D$, and $R_{th}$, which confirms the accuracy of our analytical solution presented in Theorem 1. Besides, we also observe that the SOP in all figures is no less than a constant lower bound, verifying the effectiveness of our analysis in Theorem 2. Furthermore, it can be observed from Fig. \ref{simluations} that when $P_s$ is small, increasing $P_s$ results in a lower SOP due to different scaling laws of Bob and Eve's SNRs. However, when $P_s$ is large, the SOP cannot be further improved with  increasing $P_s$, which is consistent with our analysis in Corollary 1.

\begin{figure}[!t]
	\centering
	\includegraphics[width=0.45\textwidth]{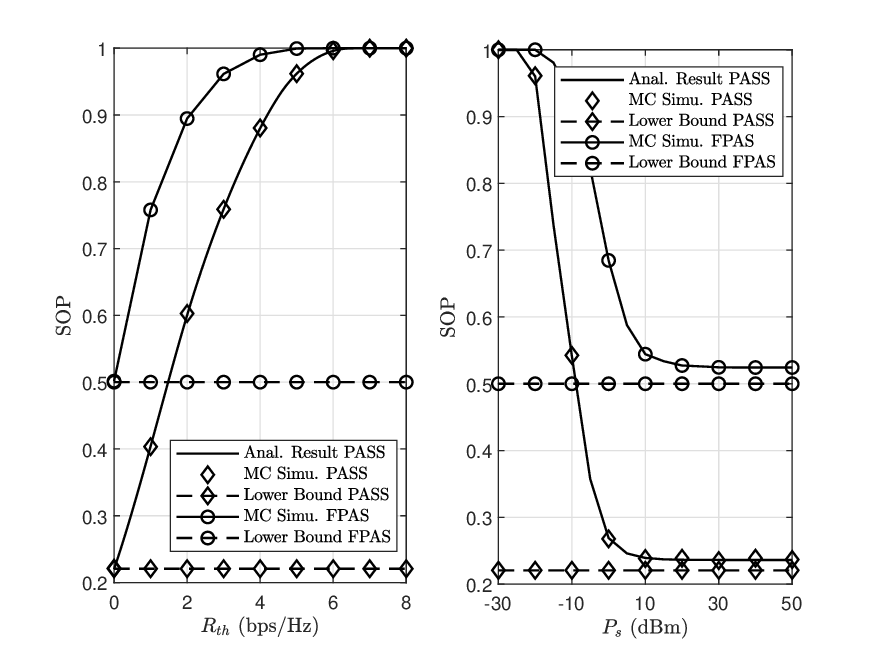} 
	\caption{Performance comparison between pinching-antenna systems (PASS) and conventional fixed-position antenna systems (FPAS).}
	\label{performance_comparison}
	\vspace{-0.3cm}
\end{figure}
In Fig. \ref{performance_comparison}, we compare the SOP of the considered pinching-antenna system with that of conventional FPA systems. The deployment region size is set to $D=30$ m. Specifically, we plot the SOP versus the target secrecy rate with $P_s=20$ dBm and the SOP versus the transmit power with $R_{th}=0.1$ bps/Hz. It can be observed that the pinching-antenna system achieves a better secrecy performance compared to the FPA system. This is attributed to the position flexibility of the pinching antenna to create a stronger line-of-sight link between the BS and Bob. Besides, the results derived from MC simulations align with both of the closed-form SOP presented in Theorem 1 and the lower-bounded SOP presented in Theorem 2.

\section{Conclusion}
In this letter, we investigated the performance of a secure pinching-antenna system. An approximate expression of the SOP was first derived in closed-form and its asymptotic behavior was analyzed in subsequence. Furthermore, we derived constant lower-bounded SOPs for the considered system and conventional FPA systems, which indicates that the former can be significantly reduced compared to the latter. Finally, our analytical results were verified via simulations, showing that the system's secrecy performance can be significantly improved compared with FPA systems due to the exceptional position flexibility of pinching antennas.

\appendices
\section{Derivation of Equation \eqref{eq7}}
Since $x_1,x_2\sim\mathcal{U}[-\frac{D}{2},\frac{D}{2}]$, it is easy to show that $x_1-x_2$ is a triangular distributed variable over $[-D,D]$. Let us define $X=x_1-x_2$. The CDF of $X^2$ can be derived as
\begin{align}
	\label{eq20}
	F_{X^2}(t)=\int_{-\sqrt{t}}^{\sqrt{t}}\frac{D-|x|}{D^2} dx=\frac{2D\sqrt{t}-t}{D^2},
\end{align}
for $0\leq t \leq D^2$. Thus, the PDF of $X^2$ is obtained as
\begin{align}
	\label{eq21}
	f_{X^2}(t)=\frac{1}{D\sqrt{t}}-\frac{1}{D^2},~0\leq t \leq D^2.
\end{align}

Then, let $Y=y_2$, with $y_2\sim\mathcal{U}[-\frac{D}{2},\frac{D}{2}]$. The PDF of $Y^2$ is given by
\begin{align}
	\label{eq22}
	f_{Y^2}(t)=\frac{1}{D\sqrt{t}},~0\leq t \leq \frac{D^2}{4}.
\end{align}

As a result, the PDF of $W=X+Y$ can be obtained as
\begin{align}
	\label{eq23}
	f_W(w)
	&=\int_{A}^{B}f_{X^2}(x)f_{Y^2}(w-x) dx \notag\\
	&=\frac{1}{D^2}\int_{A}^{B}\frac{1}{\sqrt{x(w-x)}} dx-\frac{1}{D^3}\int_{A}^{B}\frac{1}{\sqrt{w-x}} dx \notag\\
	&=\frac{1}{D^2}I_1(A,B)-\frac{1}{D^3}I_2(A,B),
\end{align}
where $A=\max(0,w-\frac{D^2}{4})$ and $B=\min(w,D^2)$. Particularly, the two integrals $I_1$ and $I_2$ can be respectively calculated as \cite{ref56}
\begin{align}
	\label{eq24}
	I_1(A,B)=2\arcsin\sqrt{\frac{B}{w}}-2\arcsin\sqrt{\frac{A}{w}}
\end{align}
and
\begin{align}
	\label{eq25}
	I_2(A,B)=2\left(\sqrt{w-A}-\sqrt{w-B}\right).
\end{align}

Then, the PDF of $\gamma_E=\frac{\bar{\gamma}}{W+h^2}$ can be obtained as
\begin{align}
	\label{eq26}
	f_{\gamma_E}(z)=\frac{\bar{\gamma}}{z^2}\cdot f_W(\frac{\bar{\gamma}}{z}-h^2).
\end{align}

Finally, by substituting \eqref{eq23} into \eqref{eq26} and after some algebraic manipulations, we can derive the closed-form PDF of $\gamma_E$, i.e., $f_{\gamma_E}(z)$, in \eqref{eq7}.

\section{Proof of Theorem 2}
For the considered pinching-antenna system, we have
\begin{align}
	\label{eq27}
	\mathcal{P}_{so}\geq \mathrm{Pr}\left(\log_2\left(1+\gamma_B\right)-\log_2\left(1+\gamma_E\right)\leq 0\right),
\end{align}

Thus, a lower bound of the SOP can be obtained as 
\begin{align}
	\label{eq28}
	\mathcal{P}_{so,low}&=\mathrm{Pr}\left(\log_2\left(1+\gamma_B\right)-\log_2\left(1+\gamma_E\right)\leq 0\right)\notag\\
	&=\mathrm{Pr}\left(\gamma_B \leq \gamma_E\right).
\end{align}

Substituting \eqref{eq3} and \eqref{eq6} into \eqref{eq28}, we can obtain
\begin{align}
	\label{eq29}
	\mathcal{P}_{so,low}=\mathrm{Pr}\left(\frac{\bar{\gamma}}{y_1^2+h^2} \leq \frac{\bar{\gamma}}{(x_1-x_2)^2+y_2^2+h^2}\right),
\end{align}
which is independent of $P_s$ and $R_{th}$. Then, let $Z=y_1$. Since $Y=y_2$ and $Z=y_1\sim\mathcal{U}[-\frac{D}{2},\frac{D}{2}]$, \eqref{eq29} can be further expanded as
\begin{align}
	\label{eq30}
	\mathcal{P}_{so,low}=&\mathrm{Pr}\left(\underbrace{(x_1-x_2)^2}_{X^2}+\underbrace{y_2^2}_{Y^2} \leq \underbrace{y_1^2}_{Z^2}\right)\notag\\
	=&8\int_{0}^{D/2}\frac{1}{D}dz\int_{0}^{z} \frac{D-x}{D^2} dx\int_{0}^{\sqrt{z^2-x^2}} \frac{1}{D} dy\notag\\
	=&\frac{8}{D^3}\int_{0}^{D/2} dz\underbrace{\int_{0}^{z} \sqrt{z^2-x^2} dx}_{J_1}\notag\\
	&-\frac{8}{D^4}\int_{0}^{D/2} dz\underbrace{\int_{0}^{z} x\sqrt{z^2-x^2} dx}_{J_2}.
\end{align}

In \eqref{eq30}, the integral term $J_1$ indicates the area of a sector with radius $z$, for which we have $J_1=\frac{\pi z^2}{4}$. Besides, defining $u=z^2-x^2$, the integral term $J_2$ can be calculated as
\begin{align}
	\label{eq31}
	J_2=\int_{u=z^2}^{u=0}-\frac{\sqrt{u}}{2} du=\frac{1}{2}\int_{0}^{z^2}\sqrt{u} du=\frac{z^3}{3}.
\end{align}

By substituting $J_1$ and $J_2$ into \eqref{eq30}, we have
\begin{align}
	\label{eq32}
	\mathcal{P}_{so,low}&=\frac{8}{D^3}\int_{0}^{D/2} \frac{\pi z^2}{4}dz-\frac{8}{D^4}\int_{0}^{D/2} \frac{z^3}{3}dz \notag\\
	&=\frac{2\pi}{D^3}\cdot\frac{(D/2)^3}{3}-\frac{8}{3D^4}\cdot\frac{(D/2)^4}{4}=\frac{2\pi-1}{24}.
\end{align} 

For conventional FPA systems, we can derive a lower bound on SOP, i.e., $\hat{\mathcal{P}}_{so,low}=0.5$, in the same way. Therefore, the proof is completed.

\bibliography{reference}
\bibliographystyle{IEEEtran}

\vfill
\end{document}